\documentstyle[aps,prl,preprint]{revtex}

\begin{document}                                                          
\titlepage
 
\title{Monte Carlo implementation of supercoiled double-stranded DNA}
\author{Zhang Yang$^{1,2}$, 
Zhou Haijun$^{1,3}$ and Ou-Yang Zhong-can$^{1,4}$}
\address{
$^1$Institute of Theoretical Physics, 
Academia Sinica, P.O. Box 2735, Beijing 100080, China\\
$^2$Institut f$\ddot u$r theoretische Physik, FU Berlin,
Arnimallee 14, 14195 Berlin, Germany\\
$^3$The State Key Lab. of Scientific and Engineering Computing,
Beijing 100080, China\\ 
$^4$Center for Advanced Study, Tsinghua University, Beijing 100084, China
}

\maketitle    

\begin{abstract}

\vskip -0.5cm
Metropolis Monte Carlo simulation is used to investigate the
elasticity of torsionally stressed double-stranded DNA, in which 
twist and supercoiling are incorporated as a natural result of
base-stacking interaction and backbone bending constrained by 
hydrogen bonds formed between DNA complementary nucleotide bases.
Three evident regimes are found in extension versus torsion and/or
force versus extension plots:
a low-force regime in which over- and underwound molecules behave
similarly under stretching;
an intermediate-force regime in which chirality appears for
negatively and positively supercoiled DNA and
extension of underwound molecule is insensitive to the supercoiling
degree of the polymer;
and a large-force regime in which plectonemic DNA is fully converted 
to extended DNA and supercoiled DNA  
behaves quite like a torsionless molecule.
The striking coincidence between theoretic calculations and recent
experimental measurement of torsionally stretched DNA
[Strick et al., Science {\bf 271}, 1835 (1996),
Biophys. J. {\bf 74}, 2016 (1998)] strongly suggests that the interplay
between base-stacking interaction and permanent hydrogen-bond constraint
takes an important role in understanding the novel properties of
elasticity of supercoiled DNA polymer. 

\end{abstract}
%\pacs{87.15.By, 36.20.Ey, 61.25.Hq, 87.10.+e}

{\Large\bf Introduction}

\medskip 
Recent years have witnessed a remarkably intense experimental and
theoretical activity in searching for the elasticity of a single
supercoiled DNA molecule (see, e.g. Strick et al., 1996, 1998; Fain et
al., 1997; 
Vologodskii and Marko, 1997;  Moroz and Nelson, 1997; 
 Bouchiat and Mezard, 1998, Zhou  et al., 1999).
Within a cell, native double-stranded DNA (dsDNA) often exists as a twisted,
and heavily coiled, closed circle. Differing amount of supercoiling,
in addition to affecting the packing of DNA within cells, influences
the activities of proteins that participate in processes --- such as
DNA replication and transcription --- that require the untwisting of
dsDNA (Wu et al., 1988).
It is believed that changes in supercoiling can also
 promote changes in DNA secondary structure that influences the
binding of proteins and other ligands (Morse and Simpson, 1988).

In recent experiments (Strick et al., 1996, 1998) 
on single torsionally constrained DNA
molecule, it was found that the supercoiling
remarkably influences the mechanical property of 
DNA molecules.
When applied with relatively low stretching force,
a supercoiled molecule can reduce
its torque by writhing, forming  structures known as
plectonemes. Therefore, the distance between two ends of
the polymer decreases with increasing supercoiling.
But above a certain critical force $f_c$, this dependence of
extension on supercoiling disappears. More strikingly, the value of
$f_c$ is significantly different for positively and
negatively supercoiled DNA,
i.e. $f_c\sim 0.8{\rm pN}$ for underwound molecule and $f_c\sim 4.5{\rm
pN}$ for overwound ones.
On the theoretical  side, 
harmonic twist elasticity and bending energy according to the wormlike
chain model have been used to understand supercoiling of DNA
polymer (Fain et al., 1997; 
Vologodskii and Marko, 1997; Moroz and Nelson, 1997;
 Bouchiat and Mezard, 1998),
and some qualitative mechanic
features of plectonemic structures of supercoiled DNA polymer
have been described by the harmonic twist elasticity.
But because of the chiral symmetry of harmonic twist elasticity,
the asymmetry of elastic behaviors of supercoiled DNA can not be
understood by this model, and especially the three obvious mechanic 
regimes observed in experiment of supercoiling DNA (Strick et al., 1996, 1998)
still need better understanding. To understand the 
supercoiling property as well as the highly extensibility  of DNA, we have
developed  a more realistic model  in which  the double-stranded nature of DNA
is taken into account explicitly   (Zhou  et al., 1999).  The supercoiling
property of {\it highly} extended DNA  was investigated analytically. 
Here, we aim at performing  a  thorough and systematic investigation into 
the property of supercoiled DNA by using 
Monte Carlo simulations based on this model.

As we have known, 
the bending energy of DNA polymer
is mainly associated with the
covalent bonding between neighboring atoms of DNA backbone
(Nossel and Lecar, 1991).
In our previous work (Zhou et al., 1999),
van der Waals interactions between adjacent basepairs was introduced
and this   helps to  explain   the highly cooperative extensibility of
overstretched DNA (Cluzel et al., 1996; Smith et al., 1996).
And it has been shown that the short-range
base-stacking interaction takes a significant role in determining the
elastical property of DNA. Lennard-Jones type potential between
adjacent basepairs can be written as  
\begin{equation}
\label{potential}
U(\theta)=
\left\{\begin{array}{cc}
\epsilon
[({\cos\theta_0\over\cos\theta})^{12}-2({cos\theta_0\over\cos\theta
})^6],
& {\rm for}\ \theta >0, \\
\epsilon [\cos^{12}\theta_0-2\cos^6\theta_0], 
& {\rm for}\ \theta\leq 0,
\end{array}
\right.
\end{equation}
(see also Fig. 1). The folding angle $\theta$ of the sugar-phosphate
backbones around DNA central axis is associated with the 
steric distance $r$ of
adjacent basepairs by $r=r_0\cos\theta$, where $r_0$ is the backbone
arclength between adjacent bases. The asymmetric
potential related to positive and negative folding angle 
$\theta$ in Fig. 1 ensures a
native DNA to take a right-handed double-helix configuration with
its equilibrium folding angle $\theta_{\rm eq}\sim\theta_0$. This
double-helix structure is  
anticipated to be very stable since $\epsilon$ ($\sim 14$ $k_B T$) is
much higher than 
thermal energy $k_BT$ according to the results of quantum chemical
calculations (Saenger, 1984).

In case that DNA polymer is torsionally  constrained, the basepair
folding angle will deviate from the equilibrium position $\theta_{\rm eq}$.
However, if the stretching force is very small, the folding angle
may deviate from $\theta_{\rm eq}$ only slightly. This is because of the following reason:
As we can infer from Fig. 1, the base-stacking potential is very sharp around $\theta_0$,
and a relatively large force is needed to make $\theta$ deviate considerably from its
equilibrium value. It is reasonable for us to anticipate that a supercoiled DNA
under low stretching force will convert its excess or deficit linking number
into positive or negative writhing of its central axis.  
Since  the central axis is symmetric with respect  to
 positive or negative writhing,
the elastic response of DNA at this force regime will certainly be symmetric with
positive or negative degree of supercoiling. Only when the stretching force 
becomes large enough will the chirality of supercoiled DNA appear. In this
regime, it becomes more and more difficult for the central axis to writhe to 
absorb linking number and  an increasing portion of the linking number will be
converted to twisting number of the backbones, which will certainly changes the
twisting manner of dsDNA. Since Eq. 1 shows that for dsDNA untwisting is 
much easier than overtwisting, chiral behavior is anticipated to emerge.
This opinion is  consistent with the experimental result of
Strick et al. (1996).

In this paper, we investigate the mechanical properties of 
supercoiled DNA by numerical Monte Carlo method. 
Base-stacking  van der Waals interactions between adjacent basepairs
are incorporated by
introducing the new degree of freedom, namely the folding angle $\theta$.
A fundamental difference from the previous approaches (See, e.g.,
Vologodskii and Marko, 1997),
which try to include the twist degrees of freedom by adding extra
terms to the free energy, is that the twist and supercoiling are
treated as the {\it  cooperative}  result of base-stacking and backbones
bending constrainted from permanent basepairs.
The striking coincidence between theoretic calculations and
experimental data of supercoiling DNA (Stick et al., 1996, 1998) 
indeed confirms this treatment. 

\bigskip
{\Large \bf Model and method of calculation}

\medskip
In the simulation, the double-stranded DNA molecule 
is modeled as a chain  of  discrete cylinders, or
two discrete wormlike chains constrained by
basepairs of fixed length $2R$ (Fig. 2). 
The conformation of 
DNA molecule of N straight cylinder segments 
is specified by the space positions of vertices of its central axis,
${\bf r}_i=(x(i),y(i),z(i))$ in 3-D Cartesian coordinate system, 
and the folding angle of the sugar-phosphate backbones
around the central axis, $\theta_i,\ i=1,2,\cdots,N$. 
Each segment is assigned the same amount of basepairs, 
$n_{\rm bp}$, so that the length of the $i$th segment satisfies 
\begin{equation}
\label{ds}
\Delta s_i=|{\bf r}_i-{\bf r}_{i-1}|=0.34 n_{\rm
bp}{\cos\theta_i\over\langle\cos\theta\rangle_0},
\end{equation} 
where $\langle\cdots\rangle_0$ means the thermal average for a relaxed
DNA molecule. Moreover, bearing in mind the experimental fact that
there are about 10.5 basepairs for each turn of a native double helix
DNA and the average distance between the adjacent basepairs is about
$d_0=0.34 {\rm nm}$, we have set 
the basepair length as $2R=(10.5d_0/\pi)\langle\tan\theta\rangle_0$ in 
our model.

Metropolis Monte Carlo method (Metropolis et al., 1953) 
is used to simulate the equilibrium evolution procedure of 
torsionally stretched dsDNA molecule. At each step of the simulation
procedure, a trial conformation of the chain is generated by
a movement from the previous one. The starting configuration
is chosen arbitrarily (except that some topology and bound
conditions should be satisfied, see below) and the averaged results of
equilibrium ensemble are independent of the initial
choice after numerous movements. 
The probability of acceptance of the movement depends on the difference in 
energy between the trial and the current conformations, according to the
Boltzmann weight.
When a trial movement is rejected, the current conformation should be
counted once more.
This procedure is repeated numerous times to obtain an ensemble of
conformations that, in principle, is representative of the equilibrium 
distribution of DNA conformation.

\bigskip
{\bf The DNA model}

\medskip

As we have known, double strand DNA is formed by winding two
polynucleotide backbones right-handedly around a common central
axis. Between the backbones nucleotide basepairs are formed with the
formation of hydrogen bonds between complementary bases. 
In our continuous model (Zhou et al., 1999),
the embeddings of
two backbones are defined by ${\bf r}_1(s)$ and ${\bf r}_2(s')$. The
ribbon structure of DNA is enforced by having ${\bf r}_2(s')$ separated from
${\bf r}_1(s)$ by a distance $2R$, i.e. 
${\bf r}_2(s')={\bf r}_1(s)+2R{\bf b(s)}$ where the hydrogen-bond-director
unit vector ${\bf b (s)}$ points from ${\bf r}_1(s)$ to ${\bf r}_2(s')$.
As the result of the wormlike backbones, the bending energy of
two backbones can be written as
\begin{equation}
\label{bending1}
E_B ={\kappa\over 2}\int_0^L[({d{\bf r}_1^2\over ds^2})^2ds+({d{\bf r}_2^2\over
	{ds'}^2})^2ds']. 
\end{equation}

The formation of basepairs
leads to rigid constraints between the two backbones and at the same
time they hinder considerably the bending freedom of DNA central axis
because of the strong steric effect. 
In the assumption of permanent
hydrogen bonds (Everaers et al., 1995; Liverpool et al., 1998; Zhou et
al, 1999), % the difference of arc-length of two backbones 
$|s'-s|=0$.
The relative sliding of backbones is prohibited
and the basepair orientation lies perpendicular to 
the tangent vectors  ${\bf t}_1={d{\bf r}_1/ds}$ and 
${\bf t}_2= {d{\bf r}_2/ ds}$ of the two backbones and 
that of the  central axis, ${\bf t}$:
${\bf b}\cdot{\bf t}_1={\bf b}\cdot{\bf t}_2={\bf b}\cdot{\bf t}=0$.
By defining the folding angle as  half of the rotation angle from
${\bf t}_2(s)$ to ${\bf t}_1(s)$, i.e., the intersection angle between
tangent vector of backbones ${\bf t}_{1(2)}$ and DNA central axis
${\bf t}$, we have
\begin{equation}
\label{permanent}
\cases{
{\bf t}_1=\cos\theta\;{\bf t}+\sin\theta\;{\bf b}\times{\bf t}\cr
{\bf t}_2=\cos\theta\;{\bf t}-\sin\theta\;{\bf b}\times{\bf t}.\cr
}
\end{equation}
Therefore, the bending energy of the two backbones can be rewritten as 
\begin{equation}
\label{bending}
E_B =
	\int_0^L[\kappa({d{\bf t}\over ds})^2+\kappa({d\theta\over
ds})^2+\kappa{\sin^4\theta\over R^2}]ds
\end{equation}
where $ds$ denotes arc-length element of the backbones, $L$ the total
contour length of each backbone, and $\kappa$ 
is the persistence length of one DNA backbone.
Bearing in mind that the pairing and stacking enthalpy of the bases
significantly increase bending stiffness of polymer axis, 
the experimental value of persistent length of dsDNA
polymer is considerably larger than that of a DNA single strand
(See, e.g. Smith et al., 1996).
To incorporate the steric effect and also considering the typical
experiment value of persistent length of dsDNA $p=53 {\rm nm}$,
the simpliest way is to substitute $k$ in the first term of
Eq. \ref{bending} with  a phenomenological parameter
$\kappa^*=53.0/2\langle\cos\theta\rangle_0 {\rm nm}k_BT$ 
 (Zhou et al., 1999), hereafter this is assumed.

Taking into account Eqs. \ref{potential} and \ref{bending}, the 
total energy of dsDNA molecule with $N$ segments in our discrete
computational model is expressed as
\begin{equation}
\label{energy}
E=
\alpha\sum_{i=1}^{N-1}\gamma_i^2+
\alpha'\sum_{i=1}^{N-1}(\theta_{i+1}-\theta_{i})^2+{\kappa\over R^2}
\sum_{i=1}^N\Delta s_i\sin^3\theta_i\tan\theta_i+ 
\sum_{j=1}^{N_{\rm bp}}U(\theta_j)-
fz(N),
\end{equation}
where $\gamma_i$ is the bending angle between the $(i-1)$th and
the $i$th segments (Fig. 2), $N_{\rm bp}$ the total number of basepairs
 of DNA
polymer, and $z(N)$ is the total extension of  the  DNA
central axis along the direction of the external force $f$ (assumed
in the $z$-direction). 

Since Kuhn statistical length of dsDNA polymer is associated with its
bending stiffness (the Kuhn
length is twice  
as persistence length of dsDNA polymer according to the wormlike chain
model), one can decide
bending rigidity parameter $\alpha$ of the discrete chain accordingly.
Suppose that we take the $N$ discrete segments to simulate the
behaviors of a dsDNA polymer of $n$ Kuhn statistical length, the length of 
$m(=N/n)$ segments should correspond to one Kuhn statistical length.
Therefore, for any chosen value $m$, 
we can decide the bending rigidity parameter $\alpha$ in
the way (see Appendix)
\begin{equation}
\label{apendix1}
m={1+\langle\cos\gamma\rangle\over 1-\langle\cos\gamma\rangle},
\end{equation}
where
\begin{equation}
\label{appendix2}
\langle\cos\gamma\rangle
={\int_0^\pi\cos\gamma\exp(-\alpha\gamma^2)\sin\gamma
d\gamma\over\int_0^\pi\exp(-\alpha\gamma^2)\sin\gamma d\gamma}.
\end{equation}

In principle, the discrete DNA model becomes continuous only when $m$
approaches infinity. 
The CPU time needed for a simulation, however, 
increases approximately as $N^2= (nm)^2$.
So it is necessary to choose a value of $m$ that is large enough to
ensure reliable results but small enough to keep the computational
time within reasonable bounds.
Our calculation and also previous work (Vologoskii et al., 1992) 
showed that
simulated properties do not depend on $m$ if it exceeds 8. Therefore,
$m=8$ was used in the current calculation, for which the bending
constant $\alpha=1.895 k_BT$.
Furthermore, we have chosen $N=160$ in consideration of the feasible
computer time. 
Since Kuhn statistical length of dsDNA is taken as $106$nm,
the B-form length of the polymer in our simulation
corresponds to $L_B=2120$nm or 6234 base-pairs. 
The constant $\alpha'$ in the second term of Eq. \ref{energy} should be
associated with stiffness of the DNA single strand. As an crude
approximation, we have taken here $\alpha'=\alpha=1.895k_BT$.\footnote{
Our unpublished data show that, the amount of second term of
Eq. \ref{energy} is quite small compared with other four terms. And
the result of simulation is not sensitive to $\alpha'$.}

The fourth term in Eq. \ref{energy} accounts for van der Waals
interactions between adjacent basepairs (see Eq. \ref{potential}). 
$\theta_0\ (=62^\circ)$ is related to the equilibrium distance between a DNA
dimer. The base-stacking intensity
$\epsilon$ is generally influenced by composition and sequence of
nucleotide chains. 
For example, the solubility experiments in biphasic systems show that
stacking interactions between purine and pyrimidine bases follow the
trend
$${\rm purine-purine} >
  {\rm pyrimidine-purine} >
{\rm pyrimidine-pyrimidine}.$$
Since we do not distinguish the specific base-sequence of purine
and pyrimidine in our DNA model, we take statistic average of stack
energies as $\epsilon =14 k_BT$,
according to the result of quantum chemical calculations (Saenger, 1984).

To simulate the extension of the stretched DNA chain, we fixed one of its ends 
at original point in 3-D Cartesian system and applied a force $f$
directed along the $z$ axis to the second end, which corresponds to
the fifth term of Eq. \ref{energy}.

\bigskip
{\bf Calculation of link number}

\medskip
The number of times the two strands of DNA double helix are interwound, 
i.e., the link number $Lk$,
is a topologic invariant quantity for closed DNA molecule and also for
linear DNA polymer in case that the orientations of two extremities of
the linear 
polymer are fixed and any part of polymer is forbidden to go round
the extremities of the polymer.
An unstressed B-DNA molecule has one right-handed twist per 3.4nm
along its length, i.e., $Lk_0=L_B/3.4$.
Under some twist stress, the link number of DNA polymer may be different
from its torsionally relexed value.
In all case when $\Delta Lk=Lk-Lk_0\neq 0$, the DNA polymer is called
``supercoiled'' (Vologodskii and Cozzarelli, 1994). 
The relative difference in link number
\begin{equation}
\label{supercoiling}
\sigma={Lk-Lk_0\over Lk_0}
\end{equation}
signifies the degree of supercoiling which is independent upon the
length of DNA polymer.
The native DNA of organisms living at physiological environment
are found always
slightly underwound and its supercoiling degree is 
between $-0.03$ and $-0.09$ (Bauer, 1978; Vologodskii and Cozzarelli,
1994), 
which is believed significantly relevant in some fundamental biological
processes (Wu et al., 1988; Morse and Simpson, 1988).

In addition to counting directly the number of times 
the two strands
are interwound, the link number of closed DNA circle
can be conveniently calculated by White's theorem (White 1969)
\begin{equation}
\label{lk}
Lk=Tw+Wr.
\end{equation}
The twist $Tw$ is the number of times basepair twist around central
axis and does not depend upon the configuration of molecule axis.
The writhe $Wr$ of molecule is a simple function of only the molecule
axis vector ${\bf r}(s)$ (White, 1969; Fuller, 1971)
\begin{equation}
\label{writhe}
Wr={1\over 4\pi}\int\int ds ds'{\partial_s{\bf
r}(s)\times\partial_{s'}{\bf r}(s')\cdot[{\bf r}(s)-{\bf
r}(s')]\over|{\bf r}(s)-{\bf r}(s')|^3}.
\end{equation}
$Wr$ is scale invariant and dimensionless and changes sign under
reflection or inversion of ${\bf r}$, reflecting the cross product in
the formula above. Therefore $Wr=0$ if ${\bf r}(s)$ is planar or
otherwise reflection symmetric.

In order to control and measure experimentally
the supercoiling degree of linear DNA polymer, 
Strick et al. (1996, 1998) attached one end of DNA molecule to a glass 
cover slip by DIG-anti-DIG links and other end to a paramagnetic bead
by biotin-streptavidin links.
Bearing in mind the diameter of magnetic bead ($\approx 4.5\mu$m)
 is far beyond that of polymer, the anchoring points can be 
considered as on impenetrable walls and $\sim 16$-$\mu$m-long DNA 
(Strick et al., 1996) in
fact is prohibited to pass around the ends of the polymer.
A magnetic field pointing
in the plane of the microscope slide was applied to fix the
orientation of the bead.
Therefore, by rotating the magnets and counting the time of turns,
the link number $Lk$ of the linear DNA molecule can be controlled and
measured experimentally. 

In Monte Carlo calculations, we restrict the DNA chain 
by two impenetrable parallel walls crossing the chain ends
which is to simulate the  above mentioned
 experimental equipment of the magnet bead and the
microscope slide (see also the treatment in Vologodskii and
Marko, 1997).
The walls are always parallel to $xy$ plane 
in our Cartian coordinate system and thus perpendicular to the
direction of the force applied to the chain ends. 

One way to calculate the link number $Lk$ of DNA molecule in our Monte
Carlo simulation is to use the White's formula Eq. \ref{lk}. However, 
the writhe $Wr$ is defined only for closed chain. In order to solve
the problem, 
we add three long flat ribbons to the two ends of the DNA chain in each
conformation during the simulation procedure. 
The axes of these ribbons are kept in the same planar 
and consist a closed circle together with the linear DNA chain. 
Since there is no any twist in the added three flat ribbons, 
it is not difficult to verify from Fig. 3 that the
number of times two strands interwind $Lk_l$ in Fig. 3a is equal to the link
number of new closed polymer $Lk_c$ in Fig. 3d.
Therefore, we only calculate $Lk$ of the closed chain in our simulations
according to Eqs. \ref{lk} and \ref{writhe}. 

Quite similar to the model by Tan and Harvey (1989) in which the twist 
of each base-pair of DNA chain is explicitly specified,
the folding angle of backbones in each segments has been
given in our model. So the twist can be directly calculated by
\begin{equation}
\label{twist}
Tw={1\over 2\pi R}\sum_{i=1}^N{\Delta s_i\tan\theta_i}.
\end{equation}
The writhe $Wr$ of the new DNA circle can be calculated
according to Eq. \ref{writhe}.

\bigskip
{\bf Simulation procedure}

\medskip
For any given force,
equilibrium sets of conformations of DNA chain
are constructed using the Metropolis MC
procedure (Metropolis et al., 1953).
Three kinds of movements have been considered in our simulations (see
Fig. 4).

In the first type of movement,
a random chosen segment is undertwisted or overtwisted by an
angle $\lambda_1$. In other words, the folding angle $\theta_i$ of
the chosen segment is modified 
into a new value $\theta'_i=\theta_i+\lambda_1$. When
$\theta'_i$ is beyond the setting interval $[-\theta_m,\theta_m]$ from 
one side, it will re-enter the interval from the opposite side
according to the periodicity assumption. 
Although the geometric limit of
folding angle of DNA backbone is $\theta_m=\pi/2$,
we set $\theta_m=85^o$ here to avoid the possible divergency in
numerical calculation of potential of Eq. \ref{potential}.
It should be mentioned that,   % as a whole configuration,
this movement modifies not 
only the folding angle of the chosen segment but also the coordinates of
all the behind vertices ${\bf r}_j,\ j=i, \cdots,N$ along the length,
since when the folding angle $\theta_i$ is changed
we have also changed the length of the segment $\Delta s_i$ according to
Eq. \ref{ds}.
So we should translate all those    segments behind this one
    to make the chain
match up (Fig. 4a).

In the second type of movement, an interval subchain containing
arbitrary amount of segments will be rotated by an angle of $\lambda_2$
around the straight line connecting the vertices bounding the
subchain (Fig. 4b). The third type of movement involves a rotation of the
subchain between a chosen vertices and the free end by an angle of
$\lambda_3$, around an axis with arbitrary orientation (Fig. 4c).
All three types of movements satisfy the basic requirement of the
Metropolis procedure of microscopic reversibility, i.e. the
probability of trial conformation $B$ when current conformation is $A$
must be equal to the probability of trial conformation $A$ when current
conformation is $B$.

All three types of movements change the configurations of DNA
chain. But from the viewpoint of energy, their functions  are
quite different. While the first type of movement concerns mainly with
modifying twist and stacking energy, the second one changes only the
bending energy and the third modifies both bending energy and extension
of DNA chain. Each of them is performed in the probability of $1/3$.
The value of $\lambda_1,\ \lambda_2,\ \lambda_3$ are uniformly
distributed over interval $(-\lambda_1^0,\lambda_1^0),\ (-\lambda_2^0, 
\lambda_2^0)$ and
$(-\lambda_3^0,\lambda_3^0)$ respectively, and $\lambda_1^0,\
\lambda_2^0$ and $\lambda_3^0$ are chosen to guarantee that about half
of the trial moves of each type are accepted.

The starting conformation of DNA chain is unknotted. But the
configurations after numerous steps of movements 
may become knotted, which violates 
the topologic invariance of chain and is incorporeal. 
Especially, both ends of molecule are anchored in the experiment
and knots never occur.
In order to avoid knotted configuration, we should check the
knot status for each trial conformation.
The most effective way to clarify the knot categories of DNA circle is to
calculate its Jones polynomial (Jones, 1985), which is strictly
topological invariant for knot categories. But the computational
calculation of Jones polynomial is quite prolix at this
moment.
In our case that it is only necessary to distinguish between unknot and
knot categories, the classical Alexander polynomial (Alexander, 1928;
Conway, 1969) is enough to meet this requirement
although it is of weaker topological invariants and 
does not distinguish mirror images.
For trivial knot, Alexander polynomial $\Delta(t)=1$;
and $\Delta (t)$ is usually not equal to $1$ for knotted chain.\footnote{
Although there are nontrivial knots whose
Alexander polynomials equal unity, this case is very rare. 
One of the example for nontrivial knot % containing 11 intersections
with $\Delta (t)=1$ can be found
in Vologodskii et al. (1974).
}
Convenient algorithms for computer calculation of Alexander polynomial 
had been well built (see, e.g. Vologodskii  et al. 1974; 
Harris and Harvey, 1999).
We only calculate the value of $\Delta (-1)$ in our simulation.
In case that the trial movement knots the chain, the energy of trial
conformation is set to be infinite, i.e. it will be rejected.

Another interaction considered in our simulation is the
steric effect of polymer chain. Since the segment has finite volume,
other segments cannot come into its own space region. This interaction
evidently swells the polymer (Doi and Edwards, 1986). 
To incorporate this exclude-volume
effect into our simulation, for each trial conformation,
we calculate the distance of between any 
point on the axis of a segment and any point on the axis of another
non-adjacent segment and check whether this distance is less than the 
DNA diameter $2R$. If the minimum distance for any two chosen segments
is less than $2R$, the energy of trial conformation is set infinite
and the movement is rejected. 

During the evolution of DNA chain, the supercoiling degree $\sigma$
may distribute around all the possible values. 
In order to avoid the waste of
computation events, we bound the supercoiling $\sigma$ of DNA chain 
inside the experimental region
(Strick et al., 1996, 1998), i.e. $-0.12\geq\sigma\geq 0.12$. 
When the torsion degree of trial conformation is beyond the chosen
range, we simply neglect the movement and reproduce a new trial movement
again. 

\bigskip
{\Large \bf Result of Monte Carlo simulation}

\medskip
To obtain equilibrium ensemble of DNA evolution, $10^7$ elementary
displacements are produced for each chosen applied force $f$.
The relative extension $x$ and supercoiling degree $\sigma$ of each
accepted conformation of DNA chain are calculated.
When the trial movement is rejected, the current conformation is 
count up twice (see Metropolis et al., 1953). 

In order to see the dependence of mechanics property of DNA upon
supercoiling degree, the whole sample is partitioned
into 15 subsamples according to the value of the supercoiling degree
$\sigma$. For each subsample, we calculate the averaged extension
\begin{equation}
\label{extension}
$$\langle x_j\rangle={1\over N_j}\sum_{i=1}^{N_j}{z_i(N)\over L_B}, \ \ \
j=1,\cdots, 15
\end{equation}
and the averaged torsion
\begin{equation}
\label{torsion}
\langle\sigma\rangle={1\over N_j}\sum_{i=1}^{N_j}\sigma_i, \ \ \
j=1,\cdots, 15,
\end{equation}
where $N_j$ is the number of movements supercoiling of which belong to
$j$th subsample.

We display the force versus relation extension for 
all positive and negative supercoiling in Fig. 5a and c respectively.
As a comparison, the experimental data (Strick et al., 1998) are shown in
Fig. 5b and d. 
In Fig. 6 is shown the averaged extension as a function of
supercoiling degree for 3 typical applied forces.
At low force, the extension in our MC simulation
saturates at a value greater than zero
because of the impenetrable walls which astrict the vertical coordinate of
the free end always higher than that of
any other points of the DNA chain. 
The same effect of the impenetrable walls was found in earlier works
(see Fig. 9 of the paper by Vologodskii and Marko, 1997).
For conciseness, we did not show the points
the relative extension of which is less than $0.15$ in Fig. 5 and 6.

In spite of quantitative difference between Monte Carlo
results and experimental data, the qualitative coincidence is
striking. Especially, three evident regimes exist in both
experimental data and our Monte Carlo simulations:
\begin{description}
\item[i).]
At a low force,
the elastic behaviour of DNA is symmetrical
under positive or negative supercoiling.
This is understandable, since the DNA torsion is the cooperative
result of hydrogen-bond constrained bending of DNA backbones and the
base-stacking interaction in our model.
At very low force, the contribution from applied force and the
thermodynamic fluctuation perturbate the folding angle $\theta$ 
of basepair to derive very little from the equilibrium position
$\theta_{0}$.
 Therefore, the DNA elasticity is achiral at this region (see  the
Introduction part of this paper).
For a fixed applied force, the increasing torsion stress tends to produce
plectonemic state which shorten the distance of two ends, therefore, the
relative extension of linear DNA polymer.
These features can be also
understood by the traditional approaches with
harmonic twist and bending elasticity (Vologodskii and Marko, 1997;
Bouchiat and Mezard, 1998).

\item[ii).]
At intermediate force, the folding angle of basepairs are pulled
slightly further away from equilibrium value $\theta_0$ where van der
Waals potential is not symmetric around $\theta_0$. So the chiral nature 
of elasticity of the DNA molecule appears. In negative supercoiling
region, i.e. $\theta < \theta_0$, the contribution of applied force
dominates that of potential because of the low plateaus of $U(\theta)$.
So the extension is insensitive to negative supercoiling degree. 
On the other hand, the positive
supercoiling still tends to contract the molecule.
\item[iii).] 
At higher force, the contribution of the applied force to the energy
dominates that of van der Waals potential in both over- and underwound
DNA. The extension of DNA accesses to its B-form length. Therefore,
the plectonemic DNA is fully converted to extended DNA, the writhe
is essentially entirely converted to twist and the force-extension
behaviour reverts to that of untwisted $(\sigma =0)$ DNA
as expected from a torsionless worm-like chain model
(Smith et al., 1992; Marko and Siggia, 1995; Zhou et al., 1999).
Because of the effect of impenetrable wall, however, the extension of DNA
molecule in our calculation is slightly higher than experimental data.
\end{description}

In conclusion, the elasticity of supercoiled double-stranded DNA is
investigated by Monte Carlo simulations.
In stead of introducing an extra twist energy term,
twist and
supercoiling  are leaded into as a nature result of cooperative
interplay between base-stacking interaction and sugar-phosphate
backbones bending constrained by permanent hydrogen bonds.
Without any adjustable parameter, the theoretic results on
the correlations among DNA extension, supercoiling degree and applied
force agree qualitatively to recent experimental data by Strick et
al (1996, 1998).

It should be mentioned that there is an up-limit of supercoiling
degree for extended DNA in current model, i.e. $\sigma_{\rm max}\sim
0.14$, which  
corresponds to $\theta =90^o$ of folding angle.
In recent experiments, Allemand et al. (1998) twisted the
plasmid up to the range of $-5<\sigma<3$. 
They found that at this ``unrealistically high'' supercoiling, the curves of
force versus extension for different $\sigma$ split again at higher
stretch force ($>3{\rm pN})$.
As argued by Allemand et al. (1998), 
in the extremely under- and overwound torsion
stress, two new DNA forms, denatured-DNA and P-DNA with exposed bases, 
will appear. 
In fact, 
if the deviation of the angle which specifies DNA twist from its
equilibrium value exceeds some threshold, the corresponding torsional
stress causes local distraction of the regular double helix structure 
(Vologodskii and Cozzarelli, 1994).
So the emergence of these two striking forms is essentially
associated with the broken processes of some basepairs under super-highly
torsional stress.
In this case, the permanent hydrogen constrain will be violated and
the configuration of base stacking interactions be varied
considerably.
We hope, with incorporation of these effects at high
supercoiling degree, our model should reproduce the novel elastic
behaviour of DNA. This part of work is in progress.

\bigskip
\bigskip
%\noindent
{\Large\bf Acknowledgements}

\medskip
Parts of the computer calculations of this work were performed 
in the Computer Cluster of Institut f$\ddot u$r
Theoretische Physik (FU-Berlin) and  the State Key Lab. of
Scientific and Engineering Computing (Beijing), which our thanks
are due to.
One of authors (Z. Y.) would like to thank
U. H. E. Hansmann, B. -L. Hao, L. -S. Liu and W. -M. Zheng 
for discussions and helps.

\bigskip
\bigskip

{\Large\bf Appendix: Kuhn Statistical Length of Discrete Chain}

\medskip
Let us consider a discrete chain of $N$ segments with each of length $l_0$,
the end-to-end vector of which is written as
\begin{equation}
\label{endtoend}
{\bf R}\equiv l_0\sum_{i=1}^N{\bf t}_i, 
\end{equation}
where ${\bf t}_i={{\bf R}_i-{\bf R}_{i-1}\over |{\bf R}_i-{\bf R}_{i-1}|}$.

For chains with bending stiffness, 
e.g. the DNA model described in Eq. \ref{energy}, 
$\langle{\bf t}_{i+k}\cdot{\bf
t}_i\rangle$ does not vanish for $k\neq 0$. 
${\bf t}_{i+k}$ can be expressed relatively to $i+k-1$'th segment as 
\begin{equation}
\label{noname}
{\bf t}_{i+k}=\cos\gamma_{i+k-1}{\bf t}_{i+k-1}+
\sin\gamma_{i+k-1}{\bf n}_{i+k-1},
\end{equation} 
where $\gamma_{i+k-1}$ is the bending angle between $i+k-1$'th and
$i+k$'th segments as defined in Eq. \ref{energy}, and
${\bf n}_{i+k-1}$ is the unit vector coplanar with ${\bf t}_{i+k}$ and
${\bf t}_{i+k-1}$ but perpendicular to the latter.
If the average of ${\bf t}_{i+k}$ is taken with the rest of the chain
(i.e., ${\bf t}_i,{\bf t}_{i+1},\cdots,{\bf t}_{i+k-1}$) fixed, one
obtains 
\begin{equation}
\label{average}
\langle{\bf t}_{i+k}\rangle_{{\bf t}_i,{\bf t}_{i+1},\cdots,{\bf
t}_{i+k-1} {\rm fixed}}=\langle\cos\gamma_{i+k-1}\rangle{\bf
t}_{i+k-1},
\end{equation}
since $\langle{\bf n}_{i+k-1}\rangle_{{\bf t}_i,{\bf t}_{i+1},\cdots,{\bf
t}_{i+k-1} {\rm fixed}}=0$ according to Eq.\ref{energy}.
Multiplying both sides of Eq. \ref{average} by ${\bf t}_i$ and taking
the average over ${\bf t}_i,{\bf t}_{i+1},\cdots,{\bf t}_{i+k-1}$, one 
has
\begin{equation}
\label{correlation}
\langle{\bf t}_{i+k}\cdot{\bf
t}_i\rangle=\langle\cos\gamma\rangle\langle{\bf t}_{i+k-1}\cdot{\bf
t}_i\rangle,
\end{equation}
where $\langle\cos\gamma\rangle$ is not specific to segments and
given by Eq. \ref{appendix2}.
This recursion equation, with the initial condition ${\bf t}^2=1$, 
is solved by
\begin{equation}
\label{correlation2}
\langle{\bf t}_{i+k}\cdot{\bf t}_i\rangle=\langle\cos\gamma\rangle^k.
\end{equation}
Thus for large $N$, $\langle{\bf R}^2\rangle$ is given by
\begin{eqnarray*}
\label{r2}
\langle{\bf R}^2\rangle
& = &
l_0^2\sum_{i=1}^N\sum_{j=1}^N\langle{\bf t}_i\cdot{\bf t}_j\rangle\\
& = &
l_0^2(N+2\sum_{i+1}^{N-1}\sum_{k=1}^{N-i}\langle{\bf t}_i\cdot{\bf
t}_{i+k}\rangle)\\
& \simeq & 
Nl_0^2{1+\langle\cos\gamma\rangle\over 1-\langle\cos\gamma\rangle}
\end{eqnarray*}

Therefore, Kuhn statistical length of the discrete chain can be
written as
\begin{equation}
\label{kuhn}
b\equiv{\langle {\bf R}^2\rangle\over R_{\rm max}}=
l_0{1+\langle\cos\gamma\rangle\over 1-\langle\cos\gamma\rangle},
\end{equation}
where $R_{\rm max}$ is the maximum length of the end-to-end vector.

\bigskip
\bigskip
\bigskip

\newpage
\noindent
{\large \bf References}
\begin{description}

%----------------------------------

\item[]
Alexander, J. W.
1928.
Topological invariants of knots and knots.
{\it Trans. Amer. Math. Soc.}, 30:275-306.

\item[] 
Allemand, J. F., D. Bensimon, R. Lavery, and V. Croquette.
1998.
Stretched and overwound DNA forms a Pauling-like structure with
exposed bases.
{\it Proc. Natl. Acad. Sci. USA} 95:14152-14157.

\item[]
Bauer, W. R.
1978.
Structure and reactions of closed duplex DNA.
{\it Annu. Rev. Biophys. Bioeng.} 7:287-313.

\item[]
Bouchiat, C., and M. Mez$\acute{a}$rd.
1998.
Elasticity model of a supercoiled DNA molecule.
{\it Phys. Rev. Lett.} 80:1556-1559.

\item[]
Cluzel, P., A. Lebrun, C. Heller, R. Lavery, J. -L. Viovy,
D. Chatenay, and F. Caron.
1996.
{\it Science}. 271:792-794.

\item[]
Conway, J. H. 
1969.
An enumeration of knots and links and some of their algebraic properties.
{\it In} Computational Problems in Abstract Algebra. 
J. Leech, editor.
Pergamon Press, Oxford.
329-358.

\item[]
Doi, M., and S. F. Edwards.
1986.
The theory of polymer dynamics.
Clarendon Press, Oxford.

\item[]
Everaers, R., R. Bundschuh and K. Kremer. 
1995.
Fluctuations and stiffness of double-stranded polymer: Railway-track
model.
{\it Europhys. Lett.} 29:263-268.

\item[]
Fain, B., J. Rudnick, and S. $\ddot{O}$stlund. 
1997. 
Conformation of Linear DNA. 
{\it Phys. Rev. E} 55:7364-7368.

\item[]
Fuller, F. B.
1971.
The writhing number of a space curve.
{\it Proc. Nat. Acad. Sci. USA.} 68:815-819.

\item[]
Harris, B. A., and S. C. Harvey.
1999.
A program for analyzing knots represented by polygonal paths.
{\it J. Comput. Chem.} 20:813-818.

\item[]
Jones, V. F. R.
1985.
A polynomial invariant for links via von Neumann algebras.
{\it Bull. Am. Math. Soc.} 12:103-112.

\item[]
Liverpool, T. B., R. Golestanian and K. Kremer. 
1998.
Statistical mechanics of double-stranded semiflexible polymers.
{\it Phys. Rev. Lett.} 80:405-408.

\item[]
Marko., J. F., and E. D. Siggia.
1995.
Strething DNA.
{\it Macromolecules} 28:8759-8770.

\item[]
Metropolis, N., A. W. Rosenbluth, M. N. Rosenbluth, and A. H. Teller.
1953.
Equation of State Calculations by Fast Computing Machines,
{\it J. Chem. Phys.} 21:1087-1092.

\item[]
Moroz, J. D., and P. Nelson. 1997. 
Torsional directed walks, entropic elasticity, and
DNA twist stiffness.
{\it Proc. Natl. Acad. Sci. USA} 94:14 418-14 422.

\item[]
Morse, R. H., and R. T. Simpson.
1988.
DNA in the nucleosome.
{\it Cell}. 54:285-287.

\item[]
Nossal, R. J., and H. Lecar, 
1991. 
Molecular and Cell Biophysics. 
Addison-Wesley Publishing Company.

\item[]
Saenger, W., 
1984. 
Principles of Nucleic Acid Structure. 
Springer-Verlag, New York.

\item[]
Smith., S. B., L. Finzi, and C. Bustamante.
1992.
Direct mechanical measurements of elasticity of single DNA molecules
by using magnetic beads.
{\it Science.} 258:1122-1126.

\item[] 
Smith, S. B., Y. Cui, and C. Bustamante,
1996.
Overstretching B-DNA: The elastic response of individual
double-stranded and single-stranded DNA molecules.
{\it Science.}  271:795-799.

\item[]
Strick, T. R., J. F. Allemand, D. Bensimon, and V. Croquette. 
1996.
The elasticity of a single supercoiled DNA molecule.
{\it Science.} 271:1835-1837.

\item[]
Strick, T. R., J. F. Allemand, D. Bensimon, and V. Croquette. 
1998.
Behavior of supercoiled DNA.
{\it Biophys. J.} 74:2016-2028.

\item[]
Tan, R. K. Z., and S. C. Harvey.
1989.
Molecular mechanics model of supercoiled DNA.
{\it J. Mol. Biol.} 205:573-591.

\item[]
Vologodskii, A. V., and N. R. Cozzarelli.
1994.
Conformational and thermodynamic properties of supercoiled DNA.
{\it Annu. Biophys. Biomol. Struct.} 23:609-643.

\item[]
Vologodskii, A. V., A. V. Lukashin, M. D. Frank-Kamenetskii and
V. V. Anshelevich. 
1974. 
The knot problem in statistical mechanics of polymer chains. 
{\it Sov. Phys. JETP} 39:1059-1063.

\item[]
Vologodskii, A. V., S. D. Levene, K. V. Klenin, M. Frank-Kamenetskii, 
and N. R. Cozzarelli. 
1992. 
Conformational and thermodynamic properties of supercoiled DNA.
{\it J. Mol. Biol.} 227:1224-1243.

\item[]
Vologodskii, A. V., and J. F. Marko.
1997.
Extension of torsionally stressed DNA by external force.
{\it Biophys. J.} 73:123-132.

\item[]
White, J. H.
1969.
Self-linking and Gauss integral in higher dimensions.
{\it Am. J. Math.} 91:693-728.

\item[]
Wu, J. H., S. Shyy, J. C. Wang, and L. F. Liu.
1988.
Transcription generates positively and negatively supercoiled domains
in the template. 
{\it Cell.} 53:433-440.

\item[]
Zhou Haijun, Zhang Yang, Ouyang Zhongcan.
1999.
Bending and Base-stacking Interactions in Double-stranded DNA.
{\it Phys. Rev. Lett.} 82:4560-4563.

%=======================================
\end{description}
 
\bigskip
\bigskip
\bigskip
%{\large\bf Figure Caption}

\begin{figure}%%%%%%%%%%%%%%%%%%%%%%%%%%%%%%%
\caption{
The van der Waals interaction potential versus folding angle of 
sugar-phosphate backbones around DNA molecule axis.
}
\label{fig1}
\end{figure}

\begin{figure}%%%%%%%%%%%%%%%%%%%%%%%%%%%%%%%
\caption{
The configuration of discrete DNA chain in our model.
}
\label{fig2}
\end{figure}

\begin{figure}%%%%%%%%%%%%%%%%%%%%%%%%%%%%%%%
\caption{
The schematic diagram to calculate link number in our simulations.
(a). For a linear supercoiled DNA chain with one end attached to a
microscope slide and with another end attached to a magnetic bead,
when the orientation of the bead is fixed and the DNA chain is forbidden 
to go round the bead, the number of times for two strands to
interwind each other, the linking number of the linear DNA ($Lk_l$), is a
topological constant.
(b). The DNA double helix is stretched to a fully extended form while the
orientation of bead keeps unchanged. The link number of linear 
DNA chain is equal to the twist number, i.e. $Lk_l=Tw_l$.
(c). Three long flat ribbons are connected to the two ends of the linear 
twisted DNA of (b). The link number of the new double helix circle is
equal to that of linear DNA chain, i.e. $Lk_c=Tw_c=Tw_l=Lk_l$ since
the writhe of the rectangle loop is 0.
(d). The DNA circle in (c) can be deformed into a new circle, 
one part of which has the same steric structure
as the linear supercoiled DNA chain in (a).
So by adding three straight ribbons,
the link number of linear double helix DNA can be obtained by
calculating the link number of the new DNA circle, 
i.e. $Lk_l=Lk_c=Tw+Wr$.
}
\label{fig3}
\end{figure}

\begin{figure}%%%%%%%%%%%%%%%%%%%%%%%%%%%%%%%
\caption{
Trial motions of the DNA chain during Monte Carlo simulations. The
current conformation of DNA central axis is shown by solid lines and
the trial conformation by dashed lines.
(a). The folding angle in $i$th segment $\theta_i$ is changed into
$\theta_i+\lambda_1$. All segments between $i$th vertex and the free
end are translated by the distance of $|\Delta s_i-\Delta s'_i|$.
(b). A portion of the chain is rotated by an angle of $\lambda_2$
around the axis connecting the two ends of rotated chain.
(c). The segments from a randomly chosen vertex to the free end are
rotated by an 
angle $\lambda_3$ around an arbitrary orientation axis which passes 
the chosen vertex. 
}
\label{fig4}
\end{figure}

\begin{figure}%%%%%%%%%%%%%%%%%%%%%%%%%%%%%%%
\caption{
Force versus relative extension curves for negatively (a,b) and positively
(c,d) supercoiling DNA molecule.
Left two plots (a) and (c) are the results of our Monte Carlo simulation,
and the horizontal bars of points denote the statistic
error of relative extension in our simulations.
Right two plots (b) and (d) are
the experimental data (Strick et al., 1998).
The solid curves serve as guides for the eye.
}
\label{fig5}
\end{figure}

\begin{figure}%%%%%%%%%%%%%%%%%%%%%%%%%%%%%%%
\caption{
Relative extension versus supercoiling degree of DNA polymer for three typical
stretch forces. Open points denote the experimental data (Strick et al., 1998) 
and solid points the results of our Monte Carlo simulation.
The vertical bars of the solid points 
signify the statistic error of the simulations,
and the horizontal ones denote the bin-width that we partition the phase
space of supercoiling degree.
The solid lines connect the solid points to guide the eye.
}
\label{fig6}
\end{figure}
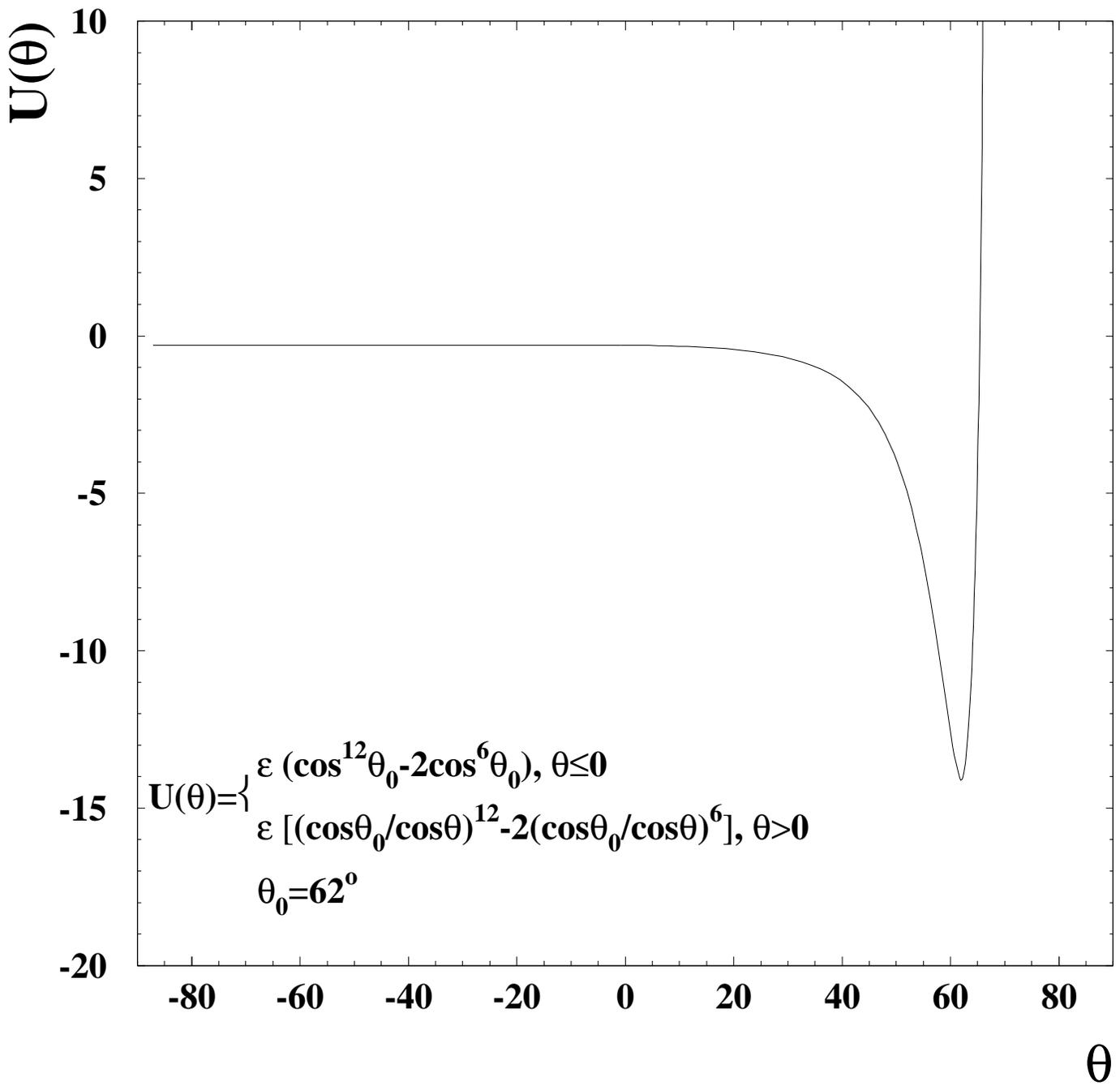
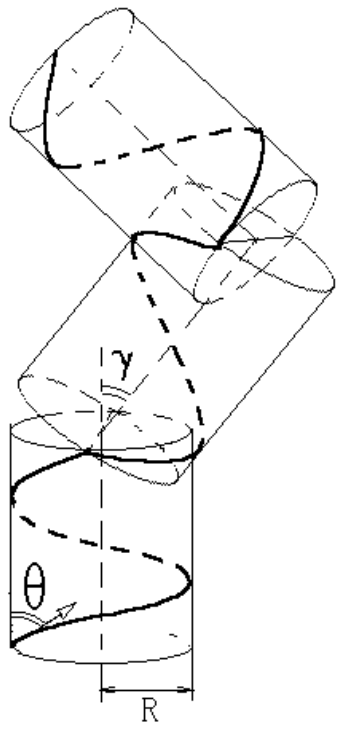
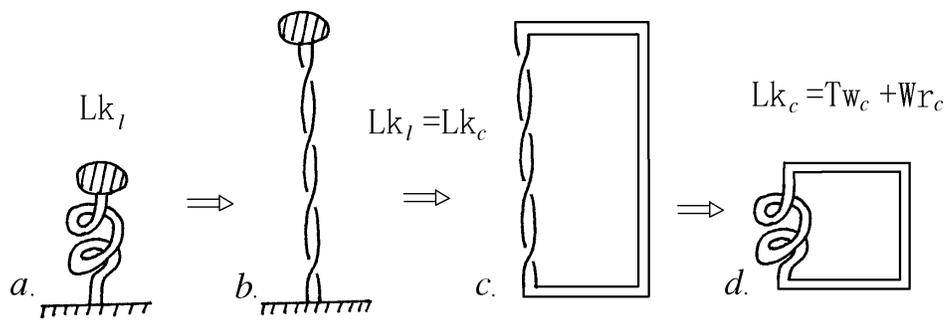
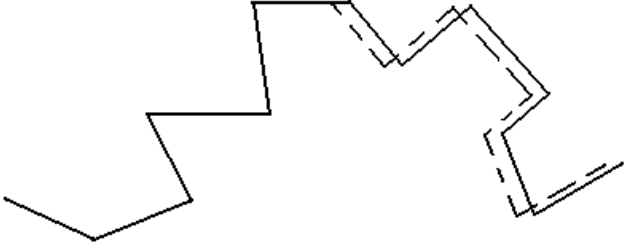
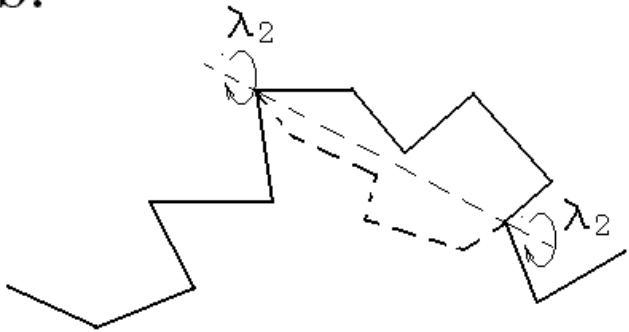
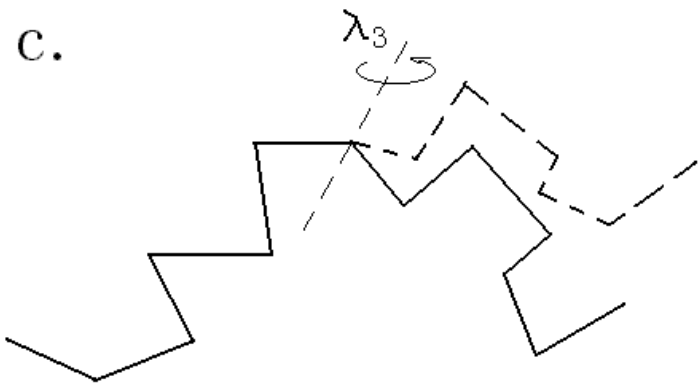
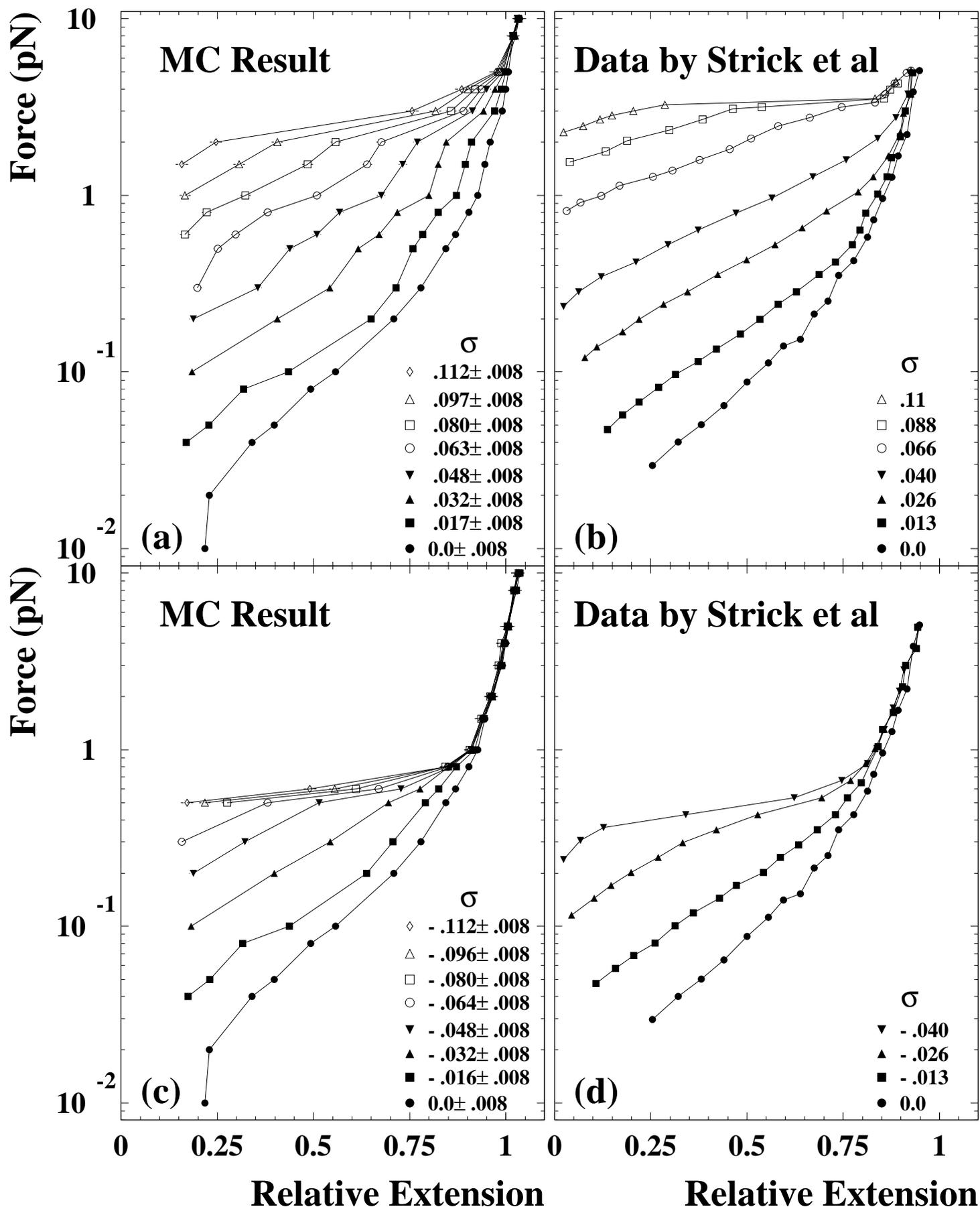
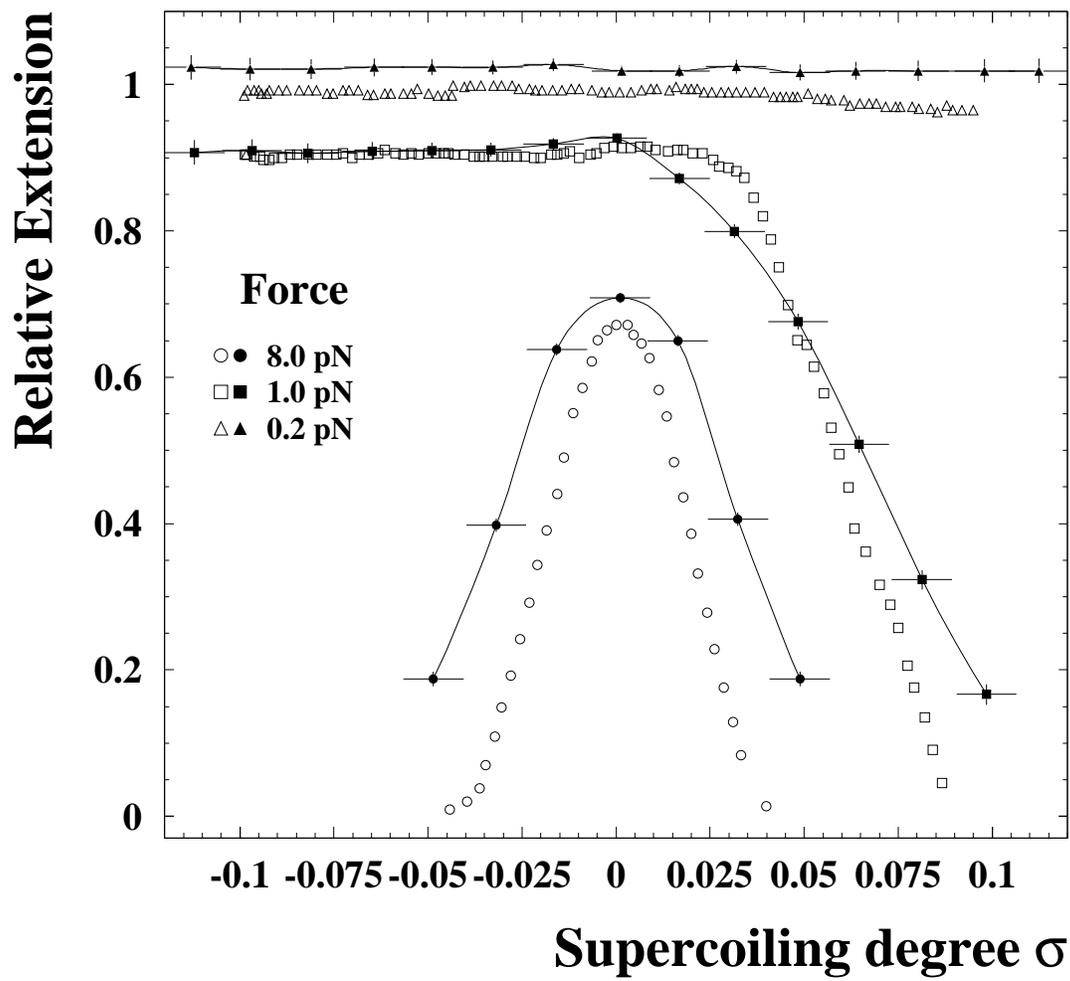
\end{document}